\documentclass[12pt]{article}

\usepackage{graphicx}
\usepackage{color,times}
\usepackage[margin=22mm]{geometry}
\usepackage{enumitem}
\usepackage{xcolor}
\usepackage[,colorlinks,urlcolor=blue,citebordercolor=blue,citecolor=blue]{hyperref}
\usepackage{isomath}
\usepackage{amsmath}
\usepackage{amsbsy}
\usepackage{amssymb}
\usepackage{amscd}
\usepackage{amsfonts}

\newcommand{\beq}{\begin{equation}}
\newcommand{\eeq}{\end{equation}}
\newcommand{\beqs}{\begin{eqnarray}}
\newcommand{\eeqs}{\end{eqnarray}}
\newcommand{\beql}{\begin{equation} \label}
\newcommand{\half}{\frac{1}{2}}

\newcommand{\comments}[1]{}

\begin{document}
\title{Anomalous phonon behavior of carbon nanotubes: First-order influence of external load}
\author{Amin Aghaei\footnote{Email: {\tt aghaei@cmu.edu}} , Kaushik Dayal\footnote{Email: {\tt kaushik@cmu.edu}} \ and Ryan S. Elliott\footnote{Email: {\tt elliott@aem.umn.edu}}
 \\ $^{* \dag}$ \small{Carnegie Mellon University, Pittsburgh, PA 15213}
 \\ $^\ddag$\small{University of Minnesota, Minneapolis, MN 55455}}

\maketitle

\begin{abstract}
External loads typically have an indirect influence on phonon curves, i.e., they influence the phonon curves by changing the state about which linearization is performed. In this paper, we show that in nanotubes, the axial load has a direct first-order influence on the long-wavelength behavior of the transverse acoustic (TA) mode. In particular, when the tube is force-free, the TA mode frequencies vary quadratically with wave number and have curvature (second derivative)
proportional to the square-root of the nanotube’s bending stiffness. When the tube has non-zero external force, the TA mode frequencies vary linearly with wave number and have slope proportional to the square-root of the axial force. Therefore, the TA phonon curves—and associated transport properties—are not material properties but rather can be directly tuned by external loads. In addition, we show that the out-of-plane shear deformation does not contribute to this mode and the unusual properties of the TA mode are exclusively due to bending. Our calculations consist of 3 parts: First, we use a linear chain of atoms as an illustrative example that can be solved in close-form; second, we use our recently developed symmetry-adapted phonon analysis method to present direct numerical evidence; and finally, we present a simple mechanical model that captures the essential physics of the geometric nonlinearity in slender nanotubes that
couples the axial load directly to the phonon curves. We also compute the density of states and show the significant effect of the external load.
\end{abstract}


\section{Introduction}
\label{sec:litrev}

The low-dimensionality of nanotubes causes unusual features in the phonon curves due to geometric effects.
In this paper, we examine particularly the transverse acoustic (TA) phonon mode and its slope and curvature (second derivative) in the long wavelength limit.

Previous works have reported four acoustic modes of vibration for SWNT.
There is consensus that two of the modes (longitudinal and twisting modes) have linear dispersion in the long wavelength limit.
However, there has been discussion about the form of the other two TA modes which are doubly degenerate.
E.g., several papers \cite{TBMD_SWCNT95, Raman_SWCNT98, Dresselhaus2000, Hone2000, Therm_prop_NT01, Maultzsch02, BN_NT02, Cao03, Symmetry_based_SWCNT03} report linear dispersion for the TA modes.
These authors do not analytically show that the TA modes should be linear, but argue that nanotubes do not have modes analogous to the out-of-plane ``bending'' mode of graphene, and thus, nanotubes must have linear long wavelength acoustic dispersion behavior.
Among these works, two recent papers \cite{BN_NT02, Symmetry_based_SWCNT03} state that quadratic dependence is found in some other papers, but argue that this may be attributed to numerical errors.

In contrast, more recent papers \cite{Ab_ini_SWNT99, Popov02, Ab_ini_SWCNT04, Lattice_model_SWCNT04, Popov06, Popov_review06, Zimmermann08, tang:wang:2011} report a quadratic dependence of frequency on the wave number.
The results presented in these papers are obtained by applying lattice dynamics methods to either tight-binding, ab-initio or empirical potential models.
In still other papers \cite{Popov2000, Mahan02, Suzuura02, Mahan04} analytical methods are used to argue that the TA mode should be quadratic.
In particular, the papers\cite{Popov2000, Mahan02, Suzuura02} make an analogy between the phonon dispersion relation and the bending energy or sound velocity in a continuous thin tube.
The paper \cite{Mahan04} develops a spring model and performs lengthy calculations for zig-zag and armchair
nanotubes\footnote{The conclusions of this study appear mixed.  In the authors' words: ``We have shown that flexure modes, those with $\omega \propto q^2$ at long wave length, exist in carbon nanotubes.  We also find that the flexure modes cross over to linear dispersion at very small values of wave  vector.  We are unsure whether this cross over is a feature of our choice of spring constants, or is an actual feature of nanotubes.''}.
Among these papers, the works\cite{Popov02, Lattice_model_SWCNT04, Zimmermann08, Suzuura02, Mahan02, Popov2000} also mention that others have observed linear dependence of the TA dispersion relations.
However, these papers do not provide a reason for the discrepancy but rather only postulate that the use of unsuitable force constant parameters may lead to linear TA phonon dispersion.

In this paper, we examine the question of the TA phonon dispersion through a combination of explicit calculations in model systems as well as numerical calculations.
Our model system consists of a linear chain of atoms as an illustrative example that can be solved in close-form.
The numerical calculations are based on recently-developed symmetry-adapted methods for nanotubes that allow the examination of bending modes\cite{dumitrica:james:2007,aghaei:dayal:2012}.
Our calculations show that in nanotubes, the axial load has a direct first-order influence on the long-wavelength behavior of the TA mode.
In particular, when the nanotube is force-free the TA mode frequencies vary quadratically with wave number and have curvature (second derivative) proportional to the square-root of the nanotube's bending stiffness.
When the nanotube has non-zero external force, the TA mode frequencies vary linearly with wave number and have  slope proportional to the square-root of the axial force.
In addition, we show that the out-of-plane shear deformation does {\em not} contribute to this mode and the unusual properties of the TA mode are exclusively due to bending.
Therefore, the TA phonon curves -- and associated transport properties -- are not material properties but rather can be directly tuned by external loads.
We illustrate this by showing the effect of the external axial load on the Density of States.
We develop a simple mechanical model based on classical beam theory that captures the essential physics of the geometric nonlinearity in slender nanotubes that couples the axial load directly to the phonon curves.

The paper is organized as follows.
We first describe what is known in layered nanostructures in Section \ref{layered}.
We then present the closed-form calculations for atomic chains in Section \ref{chains}.
In Section \ref{SWCNT}, we present the results of numerical calculations for carbon nanotubes.
Finally, in Section \ref{beam}, we show that a simple model based on classical beam theory captures the key nonlinearity that governs the TA mode behavior.

%


\section{Layered crystals and atomic sheet structures}
\label{layered}

Quadratic dependence on the wave vector for long wavelength TA phonons is well
known in the context of layered crystals, such as graphite, and sheet
structures, such as graphene.  The first works in this area were by
Komatsu\cite{Komatsu51, Komatsu55} (1951) and Lifshitz\cite{Lifshitz52} (1952).
Their work showed that the TA phonon mode, with wave vector parallel to the
layers, is a ``bending mode'' and has quadratic dispersion.  Other works
concerned with bending modes in layered crystals include \cite{Komatsu58,
  Therm_expan_graphi70, Neg_therm_expan93, Zabel01, Savinia11}.  The dispersion
relation results of these early investigations are summarized in the
paper\cite{Therm_expan_graphi70} as\footnote{Note, these equations correspond
  to the results of Lifshitz's work.  Komatsu gives similar expressions except
  that he replaces the coefficients of the wave vector components by the
  squared sound velocities of graphite.}
\begin{align} 
  \text{Transverse in-plane mode: } & \quad
  \omega_1^2 = \frac{C_{11}-C_{12}}{2\rho} \left( k_x^2 + k_y^2 \right)
  + \frac{C_{44}}{\rho} k_z^2,  \label{Lifshitz1} \\
  \text{Longitudinal in-plane mode: }  & \quad
  \omega_2^2 = \frac{C_{11}}{\rho} \left( k_x^2 + k_y^2 \right) +
  \frac{C_{44}}{\rho} k_z^2,
  \label{Lifshitz2} \\
  \text{Out-of-plane (bending) mode: } & \quad
  \omega_3^2 = \frac{C_{44}}{\rho} \left( k_x^2 + k_y^2 \right)
  + \frac{C_{33}}{\rho} k_z^2 + \Lambda^2 \left( k_x^2 + k_y^2
  \right)^2.  \label{Lifshitz3}
\end{align}
Here, $C_{11}, C_{12}, C_{33}$ and $C_{44}$ are the elastic constants of the
crystal, $\rho$ is the density, $k_x$ and $k_y$ are the components of the wave
vector parallel to the layer planes and $k_z$ is the component of the wave
vector perpendicular to the layers.  The quantity $\Lambda$ characterizes the
\emph{bending rigidity} of layers and is determined mainly by intralayer
forces.  For layered crystals, the interlayer elastic constants $C_{33}$ and
$C_{44}$ are usually negligible compared to $\Lambda$.  This illustrates that
the predominant contribution to the bending mode dispersion relation is due to
the bending rigidity $\Lambda$.

It is important to note that Eqn. \eqref{Lifshitz3} holds for stress-free
layers only.  If the layer or sheet is subjected to non-zero in-plane stress
then, according to Lifshitz, the dispersion of the bending mode is
\begin{equation}  \label{Lifshitz4}
  \omega_3^2 = \frac{\sigma}{\rho} \left( k_x^2 + k_y^2 \right)
\end{equation}
where $\sigma$ is a measure of the stress in the layer.  Thus, the TA bending
mode dispersion relation of a stressed layer has a linear dependence on the
long wavelength wave vector and has a slope proportional to the square-root of
the stress.  Making an analogy between layered crystals or sheets and
nanotubes, one would then expect similar behavior of the TA phonon modes in
nanotubes.  In fact, this is exactly what we find in Section~\ref{SWCNT}
from our numerical calculations.



\section{Phonons in a Linear Atomic Chain}
\label{chains}

In this section, we examine a linear atomic chain as a model system that permits closed-form calculations.
We show that the TA modes of a linear chain of atoms have linear dependence on the wave number when the chain is subjected to a non-zero axial force.
Further, we show that the TA long wavelength phonon modes for a \emph{force-free} chain correspond to bending, not shear.


\subsection{Transverse acoustic phonon modes}
For simplicity, we consider a linear chain of atoms which is free to deform in
a two-dimensional space.  In the configuration of interest the position vector
of atom $j$ is $\mathbf{x}_j = x_j^\alpha \mathbf{e}_\alpha = a j
\mathbf{e}_1$, where $a$ is the lattice spacing of the chain, $\mathbf{e}_1$ is
the axial direction of the chain, and $\alpha \in \{1,2\}$.  A standard
argument leads to the eigenvalue equation for the phonon frequencies
\begin{equation}
  \mathbf{D}_k \mathbf{v}_k = m \omega_k^2 \mathbf{v}_k,
\end{equation}
where $m$ is the mass of each atom in the chain, $k$ is the wave number (along
the axial direction), $\omega_k$ is the phonon frequency, $\mathbf{v}_k$ is the
phonon polarization vector (eigenvector), and $\mathbf{D}_k$ is the dynamical
matrix given by
\begin{equation}
  \mathbf{D}_k = \sum_j \mathbf{H}_{0j} \exp[-i k x_j^1].
\end{equation}
Here, $\mathbf{H}_{pq}$ is the Hessian of the chain's potential energy function
$\phi = \phi(\{\mathbf{x}_j\})$.  That is, $\mathbf{H}_{pq}
= \partial^2\phi/\partial\mathbf{x}_p\partial\mathbf{x}_q$.  Due to the
Euclidean invariance (objectivity) of the potential energy, we can---without
loss of generality---take $\phi = \bar{\phi}(\{r_{pq}\})$ to be a function of
the pair-distances ($r_{pq} \equiv \| \mathbf{x}_q - \mathbf{x}_p \|$) between
atoms in the chain\cite{tadmor:miller:2011}.  Then we find
\begin{equation}
  \begin{aligned}
    \mathbf{H}_{0q} = &\sum_{i\not=0} \sum_{j\not=q} \frac{\partial^2 \bar{\phi}}{\partial r_{0i} \partial r_{qj}} \frac{\partial
      r_{0i}}{\partial \mathbf{x}_0} \frac{\partial r_{qj}}{\partial
      \mathbf{x}_q} + \sum_{i\not=0} \frac{\partial \bar{\phi}}{\partial
      r_{0i}} \frac{\partial^2 r_{0i}}{\partial \mathbf{x}_0 \partial
      \mathbf{x}_q}\\
    = \sum_{i\not=0} &\sum_{j\not=q} \frac{\partial^2 \bar{\phi}}{\partial
      r_{0i} \partial r_{qj}} \frac{1}{r_{0i} r_{qj}} (\mathbf{x}_i -
    \mathbf{x}_0)\otimes(\mathbf{x}_j - \mathbf{x}_q)\\ &+ \delta_{0q}
    \sum_{i\not=0} \frac{\partial \bar{\phi}}{\partial r_{0i}}
    \frac{1}{r_{0i}^3} \left[ r_{0i}^2 \mathbf{I} - (\mathbf{x}_i -
      \mathbf{x}_0)\otimes(\mathbf{x}_i - \mathbf{x}_0)\right]\\ &-
    (1-\delta_{0q})\frac{\partial \bar{\phi}}{\partial r_{0q}}
    \frac{1}{r_{0q}^3} \left[ r_{0q}^2 \mathbf{I} - (\mathbf{x}_q -
      \mathbf{x}_0)\otimes(\mathbf{x}_q - \mathbf{x}_0)\right].
  \end{aligned}
\end{equation}
In the standard basis, this becomes
\begin{equation}
  [H_{0q}]^{\alpha\beta} = \begin{bmatrix}
    \sum_{i\not=0}\sum_{j\not=q} \frac{\partial^2
      \bar{\phi}}{\partial r_{0i} \partial r_{qj}} \frac{i}{|i|}\frac{j-q}{|j-q|}&
    0\\
    0 &
    \delta_{0q} \sum_{i\not=0} \frac{\partial \bar{\phi}}{\partial r_{0i}}
    \frac{1}{|i| a} - (1-\delta_{0q}) \frac{\partial \bar{\phi}}{\partial r_{0q}}
    \frac{1}{|q| a}
  \end{bmatrix}.
\end{equation}

From this form of the Hessian, it is clear that the polarization vector for the
TA phonon modes of the linear atomic chain is $\mathbf{v}_{\text{TA}} = [0,
1]^{\text{T}}$.  Thus for this case, the TA dispersion relation is found to be
\begin{equation}
  \begin{aligned}
    \omega_{\text{TA},k}^2 &= \frac{1}{m} \sum_{j} H_{0j}^{22} \exp\left[-i
      k j a\right]\\
    &= \frac{1}{m} \sum_{j\not=0} \frac{\partial \bar{\phi}}{\partial
        r_{0j}} \frac{1}{|j| a} (1 - \exp{\left[-i k j a\right]}).
  \end{aligned}
\end{equation}
Expanding the exponential in powers of (the small parameter) $k$ and noting
that $\partial\bar{\phi}/\partial r_{0j} = \partial\bar{\phi}/\partial
r_{0-j}$, results in
\begin{equation}
  \omega_{\text{TA},k}^2 = \left[ \frac{1}{2 m} \sum_{j\not=0} \frac{\partial
      \bar{\phi}}{\partial r_{0j}} |j|a \right] k^2 + \left[ \sum_{j\not=0}\frac{-1}{24 m}
    \frac{\partial \bar{\phi}}{\partial r_{0j}} |j|^3 a^3 \right] k^4 + \dots.
\end{equation}
Noting that $|j|a = r_{0j}$ and neglecting higher order terms we find
\begin{equation}\label{eq:chain:TAdisp}
  \omega_{\text{TA},k}^2 = \left[ \frac{1}{2 m} \sum_{j\not=0} \frac{\partial
      \bar{\phi}}{\partial r_{0j}} r_{0j} \right] k^2 + \left[ \frac{-1}{24 m}
    \sum_{j\not=0} \frac{\partial \bar{\phi}}{\partial r_{0j}} r_{0j}^3\right] k^4.
\end{equation}

The virial axial force in a chain of atoms at zero temperature can be shown \cite{aghaei2011symmetry,aghaei2012tension} to
be $F=\frac{1}{2L} \sum_i \sum_j \frac{\partial \phi}{\partial r_{ij}} r_{ij}
=\frac{1}{2a} \sum_j \frac{\partial \phi}{\partial r_{0j}} r_{0j}$.
Thus, the
linear part of the dispersion relation ($\omega$ versus $k$) is proportional to
the square-root of the axial force in the chain.  Next, we show that the second
term in \eqref{eq:chain:TAdisp} is related to the bending stiffness of the
chain.


\subsection{Energy change corresponding to shear deformation}
If $\gamma$ is the shear parameter in the infinite chain, the displacement of
the atoms is $u^1_j = 0$ and $u^2_j=\gamma x^1_j$. The separation between the
atoms in the deformed chain will be
\begin{equation}
  r_{jl}+\Delta r_{jl} = \left[ r_{jl}^2 + (u^2_l - u^2_j)^2 \right]^{1/2}
  = \left[ r_{jl}^2 + \gamma ^2 r_{jl}^2 \right]^{1/2}
  = r_{jl} \left[ 1 + \gamma^2 \right]^{1/2}
  \approx r_{jl} \left[ 1 + \frac{\gamma^2}{2} + \frac{-\gamma^4}{4} +\cdots
  \right].
\end{equation}
Therefore the change in distance between the atoms is $\Delta r_{jl} \approx
\frac{\gamma^2}{2} r_{jl} = \frac{\gamma^2}{2} r_{0(j-l)}$. Expanding the
potential energy density, $\tilde\phi=\bar\phi/2Na$, up to the first order in
perturbation, gives an energy change of the form
\begin{equation}
  \Delta \tilde\phi \approx
  \frac{1}{2Na} \sum_{j,l\ne j} \frac{\partial \bar\phi}{\partial r_{jl}}
  \Delta r_{jl} = \frac{\gamma^2}{4Na} \sum_{j,l\ne j} \frac{\partial
    \bar\phi}{\partial r_{0(l-j)}} r_{0(j-l)} = \frac{\gamma^2}{2a} \sum_{n
    \ne 0} \frac{\partial \bar\phi}{\partial r_{0n}} r_{0n} = F \gamma^2,
\end{equation}
where $F$ is the virial axial force. As can be seen, the change in energy due
to shear deformation is not related to the second term of
\eqref{eq:chain:TAdisp}.

\subsection{Energy change corresponding to bending deformation}
Now consider the chain subjected to bending. The atom with initial position
$\mathbf{x}_j=ja\mathbf{e}_1$ will move to $\mathbf{y}_j= r [\sin (aj/r)
\mathbf{e}_1 + \cos (aj/r) \mathbf{e}_2]$, where $r$ is the radius of
curvature.  The arc length does not change during the bending.  If $r_{jl} =
(j-l)a$ is the separation between atoms $j$ and $l$ before the bending, after
bending the separation will be
\begin{align}
  r_{jl}+ \Delta r_{jl}  &= \left[ (y^1_l - y^1_j)^2 + (y^2_l - y^2_j)^2 \right]^{1/2}
  = \left[ r^2 \left(\cos \frac{al}{r} - \cos \frac{aj}{r} \right)^2
    + r^2 \left(\sin \frac{al}{r} - \sin \frac{aj}{r} \right)^2 \right]^{1/2}  \notag \\
  &= \sqrt{2} r \left[ 1- \cos \frac{al}{r} \cos \frac{aj}{r} - \sin \frac{al}{r}
    \sin \frac{aj}{r} \right]^{1/2} = \sqrt{2} r \left[ 1- \cos
    \frac{a(l-j)}{r} \right]^{1/2} \notag \\
  &= 2r \sin  \frac{r_{jl}}{2r}   \notag  \\
  & \approx r_{jl} - \frac{r_{jl}^3}{24r^2}.
\end{align}
This shows that 
\begin{equation}
  \Delta r_{jl} \approx - \frac{r_{jl}^3}{24r^2}.
\end{equation}
Expanding the potential energy density up to the first order in perturbation
gives an energy change of the form
\begin{equation}
  \Delta \tilde\phi= 
  \frac{1}{2Na} \sum_{j,l\ne j} \frac{\partial \bar\phi}{\partial r_{jl}}
  \Delta r_{jl} 
  = \frac{-1}{48Nr^2 a} \sum_{j,l\ne j} \frac{\partial \bar\phi}{\partial
    r_{0(l-j)}} r_{0(j-l)}^3
  = \frac{-1}{24r^2 a} \sum_{n  \ne 0} \frac{\partial \bar\phi}{\partial
    r_{0n}} r_{0n}^3. 
\end{equation}
Hence, we find $\Delta \tilde\phi \approx \Lambda \frac{1}{r^2} = \Lambda
\kappa^2$, where
\begin{equation}
  \Lambda = \frac{-1}{24a} \sum_{n \ne 0} \frac{\partial \bar\phi}{\partial
    r_{0n}} r_{0n}^3
\end{equation}
is the bending stiffness and $\kappa=1/r$ is the bending curvature.  Thus, we
have shown that the second term on the right hand side of
\eqref{eq:chain:TAdisp} is related to the bending stiffness $\Lambda$.
Consequently, the curvature of the force-free chain's TA dispersion relation is
proportional to the square-root of its bending stiffness.


\section{Numerical Calculations for Carbon Nanotubes}
\label{SWCNT}

\begin{figure}
	\centering
	\includegraphics[scale=0.2, clip]{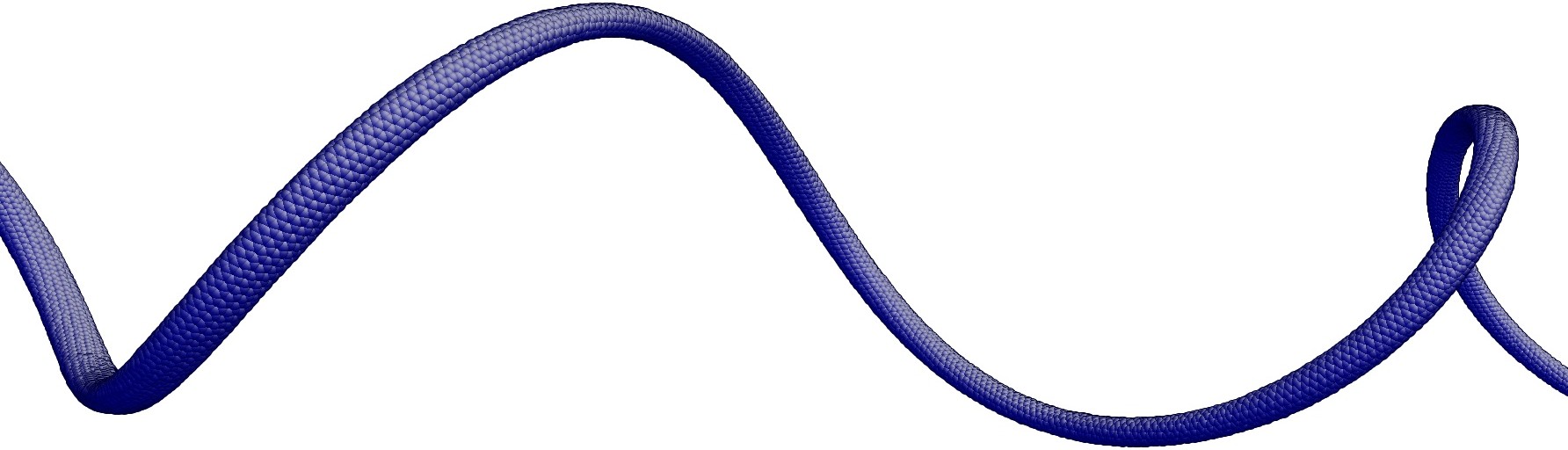}
	\caption{The distorted shape, i.e., the eigenmode, of the nanotube due to the transverse acoustic mode, near the long-wavelength limit of a (7, 6) nanotube.}
	\label{fig1}
\end{figure}

In this section, we use our recently developed symmetry-adapted phonon analysis method\cite{aghaei:dayal:2012} to numerically demonstrate the connection between the TA dispersion relation and the axial force and bending stiffness of a SWNT. The distorted shape, i.e., the eigenmode, of the nanotube due to the transverse acoustic mode, near the long-wavelength limit of a (7, 6) nanotube is plotted in Fig. \ref{fig1}. All calculations are performed using the Tersoff interatomic potential \cite{tersoff:1988}.

Before we describe our results, we note that the paper
\cite{Continuum_nanowire97} shows, numerically, that bending modes with
quadratic dispersion exist in quantum wires with rectangular cross section.
The existence of only one continuous wave vector makes the wires similar to
nanotubes.  However, contrary to nanotubes, because of the rectangular cross
section, the bending modes are no longer degenerate in wires.  Our systematic
symmetry-adapted phonon analysis method is able to deal with the degenerate
behavior that occurs in nanotubes and clearly illustrates the quadratic
dispersion of the TA phonon modes for a force-free tube.

First, we investigate the effect of axial relaxation on the density of states
(DOS) and the form of the dispersion relation for the bending modes. Then we
numerically show that the slope and curvature of the TA modes correspond to the
axial force and bending rigidity, respectively.

\begin{figure}
	\centering
	\includegraphics[scale=0.7, trim=7mm 95mm 5mm 90mm, clip]{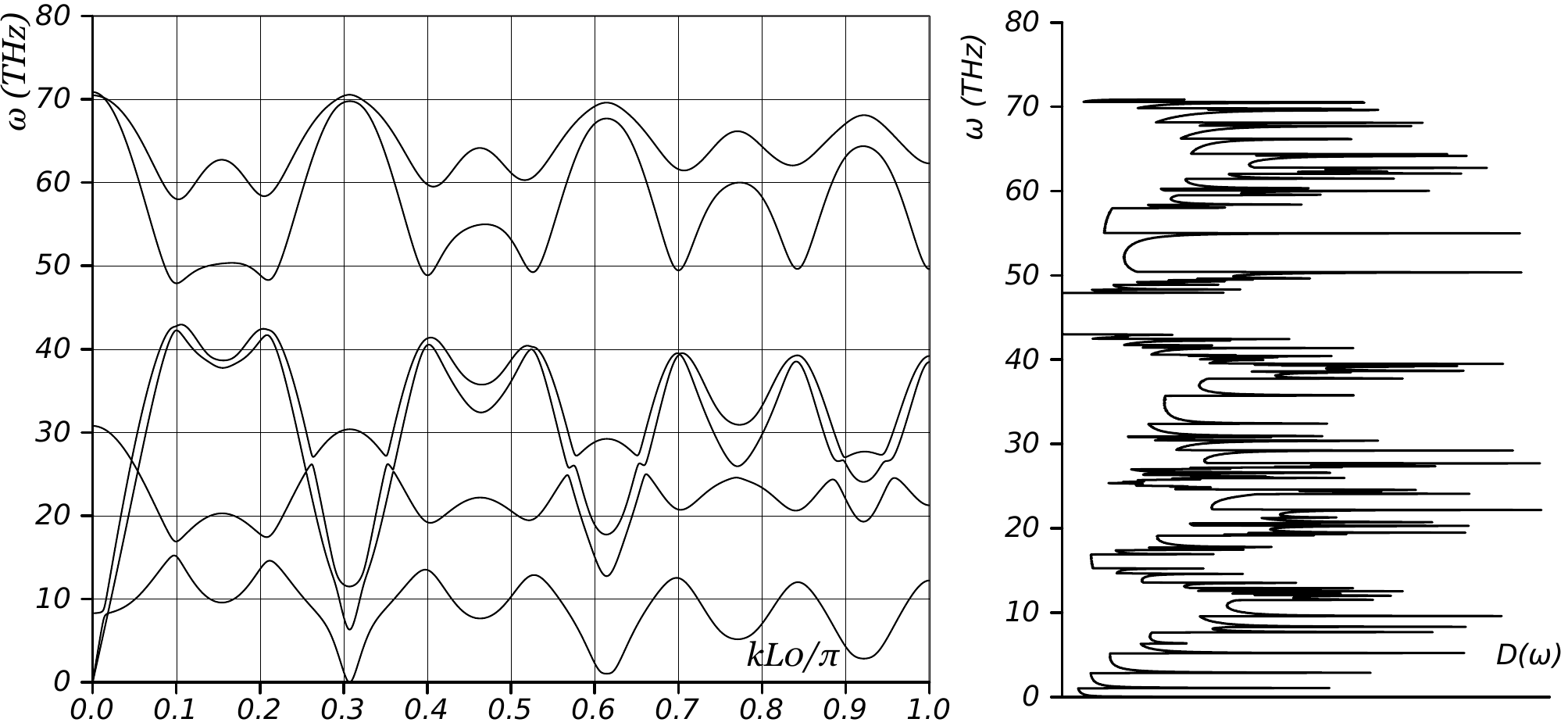}
	\caption{(Left) Dispersion curves and (right) DOS of zero-force axially
          relaxed (7,6) SWNT as described within the symmetry-adapted objective
          structures framework of paper\cite{aghaei:dayal:2012}}
	\label{unstrained}
\end{figure}

\begin{figure}
	\centering
	\includegraphics[scale=0.5]{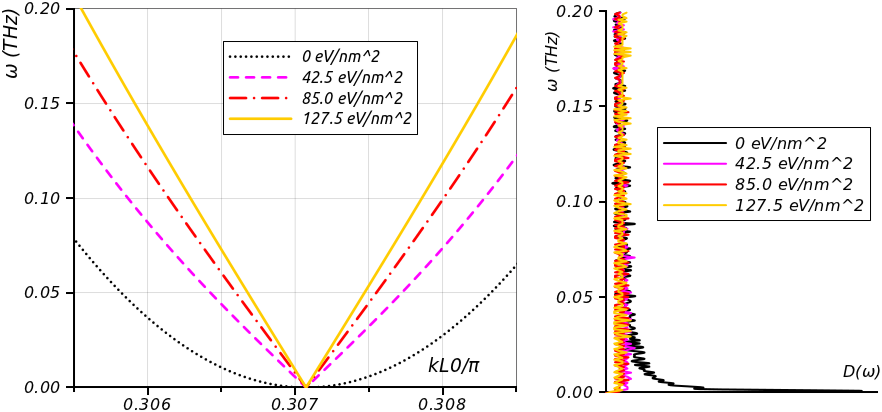}
	\caption{The effect of axial stress on (left) the dispersion relation
          of the TA phonon modes and (right) the DOS of a (7,6) SWNT}
	\label{strained}
\end{figure}

Following the procedures outlined in our previous
paper\cite{aghaei:dayal:2012}, the dispersion curves and DOS for a relaxed
(force-free) (7,6) SWNT are plotted in Fig.~\ref{unstrained}.  Adding axial
force leads to one significant change of the dispersion curves and DOS.
That is, the dispersion of low frequency TA modes changes from quadratic to
linear in the long wavelength limit (as $k\rightarrow \theta$), where $\theta =
0.307\pi$ is the rotation angle of the $(7,6)$ nanotube screw generator.  A
consequence of the low frequency dispersion curve near $k=\theta$ is that the
DOS is singular near $\omega = 0$ for the force-free tube, while conversely, it
is almost constant for tubes with non-zero force.  This effect is illustrated
in Fig.~\ref{strained}, by plotting the long wavelength TA phonon
dispersion relation near $k=\theta$ for tubes subjected to four different
values of axial stress.  Here, we define the axial stress as the axial force
divided by the force-free circumference of the tube.

The figure shows clear evidence that the long wavelength TA phonon dispersion
behavior is related to the axial stress in the SWNT.

To show more conclusively that the slope and curvature of the long wavelength
TA modes have the specific dependence postulated in this paper, we compute (1)
the bending modulus of SWNTs as a function of tube diameter, (2) the slope of
the long wavelength TA phonon mode dispersion relation as a function of axial
stress and tube diameter, and (3) the curvature of the long wavelength TA
phonon mode dispersion relations for a force-free SWNT as a function tube
diameter.

\begin{figure}
	\centering
	\includegraphics[scale=0.35, clip]{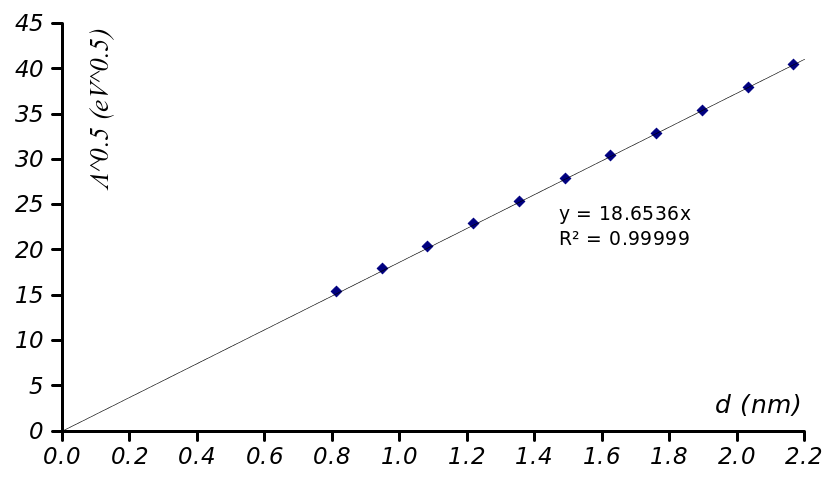}
	\caption{The square root of bending stiffness (calculated by minimizing
          the energy of armchair nanotubes) versus the diameter of the
          nanotube.  The points are from calculations and the line is a best-fit to the points.}
	\label{MD}
\end{figure}

We compute the energy density versus curvature relation using zero-temperature
objective molecular dynamics (based on \cite{dumitrica:james:2007} which allows for the explicit control of
curvature through a ``bending group parameter'') for armchair nanotubes with
different diameters.  In particular, we calculate the change in energy density,
$\Delta\tilde{\phi}= \Delta\bar{\phi}/(\ell P)$, where $\ell$ is the objective
cell length and $P$ is the perimeter of the nanotube.  For each radius, we
calculate the bending stiffness $\Lambda$ according to
\begin{equation}
  \Delta\tilde{\phi} = \half \Lambda \kappa^2  \quad  \Rightarrow  \quad
  \Lambda=\frac{2 \Delta\tilde{\phi}}{\kappa^2}.
\end{equation}
Figure~\ref{MD} shows the variation of $\sqrt \Lambda$ versus the diameter.
It is clear that the bending stiffness scales linearly with the square of the
diameter.

\begin{figure}
	\centering
	\includegraphics[width=0.5\textwidth]{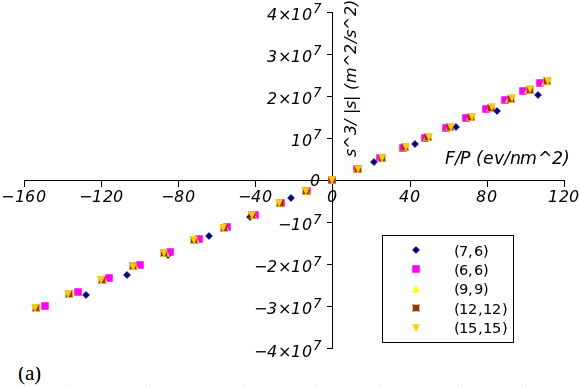}%
	\includegraphics[width=0.5\textwidth]{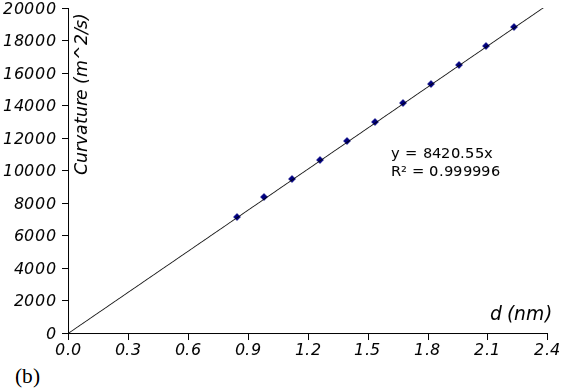}
	\caption{(a) Square of the slope of the TA mode vs.\ the axial
          stress. (b) Curvature of the TA mode vs.\ diameter for various nanotubes ranging from $(6,6)$ to $(16,16)$.  The points are from calculations and the line is a best-fit to the points.}
	\label{Phonon}
\end{figure}

Next, Fig.~\ref{Phonon}(a) shows the (signed) square of the TA modes' slope
(near $k=\theta$) as a function of axial stress for a variety of tube SWNT
types.
This slope was computed from our phonon dispersion calculations using a
numerical differentiation algorithm.  Finally, in Fig.~\ref{Phonon}(b) we
show the curvature, computed in a similar manner to the slope, for various nanotubes ranging from $(6,6)$ to $(16,16)$
 as a function of the tube diameter.

From these figures we can infer the following important results:
\begin{itemize}
	\item Fig. \ref{Phonon}(a) shows that the slope of the TA mode curve does not depend on the specific nanotube.  It is linearly related to the square root of the axial force.
	\item For various nanotubes of different diamters and chiralities, Fig. \ref{Phonon}(b) shows that the curvature of the TA mode curve scales linearly with the tube diameter, and Fig. \ref{MD} in turn shows that square root of the bending stiffness scales linearly with the tube diameter.  Together, these imply that the curvature of the TA mode curve is directly related to the square root of the bending stiffness.
\end{itemize}

These plots therefore confirm our claim.

\section{Simple Continuum Model and Discussion}
\label{beam}

First we consider the case of a rod that undergoes stretching deformations, but with a pre-stress.
This gives us the expected conclusion that the pre-stress plays a role only in determining the equilibrium state around which linearization is conducted.
Consider a rod of length $L$ that is extend by an amount $\delta$ and with a pre-stress $\sigma_0$.
The energy of deformation is, to second order, given by
\begin{equation}
	E \sim \sigma_0 \frac{\delta}{L} + \half E_T \left( \frac{\delta}{L} \right)^2 + \ldots
\end{equation}
Here $E_T$ is the extensional modulus at the state of linearization.
The natural frequencies from a normal mode analysis will be proportional to the square root of the coefficient of the quadratic term, i.e. $E_T^\half$.

Now, to understand the current setting, we examine a beam that undergoes transverse deflections.
Consider a beam of length $L$, and impose a bending deformation that causes a relative deflection $\Delta$ between the ends of the beam.
A simple accounting of the geometric nonlinearity shows that there is an {\em extensional} strain of $\left( 1+\left(\frac{\Delta}{L}\right)^2 \right)^\half - 1$ which is simply $\half \left(\frac{\Delta}{L}\right)^2$ to leading order.
The energy of deformation is, to second order, given by
\begin{equation}
	E \sim  \half \sigma_0 \left( \frac{\Delta}{L} \right)^2 + \ldots
\end{equation}
The natural frequencies from a normal mode analysis will be proportional to the square root of the coefficient of the quadratic term, i.e. $\sigma_0^\half$.
When this term is absent, higher-order effects such as bending control the curvature of the TA mode curve.

Therefore, this gives the unusual {\em direct} dependence of the normal mode frequencies on the axial force.
As the above simple analysis shows, the key to this feature is the geometric nonlinearity of a slender structure.

As our calculations above have shown, the Density of States can also be tuned significantly by applying an axial force.
Therefore, it is possible that properties such as heat transfer can be tuned by mechanical means.

\section*{Acknowledgments}

Amin Aghaei and Kaushik Dayal thank AFOSR Computational Mathematics (FA9550-09-1-0393) and AFOSR Young Investigator Program (FA9550-12-1-0350) for financial support.
Kaushik Dayal also acknowledges support from NSF Dynamical Systems (0926579), NSF Mechanics of Materials (CAREER-1150002) and ARO Solid Mechanics (W911NF-10-1-0140).
Amin Aghaei also acknowledges support from the Dowd Graduate Fellowship.
This work was also supported in part by the NSF through TeraGrid resources provided by Pittsburgh Supercomputing Center.
Kaushik Dayal thanks the Hausdorff Research Institute for Mathematics at the University of Bonn for hospitality.
We thank Richard D. James for useful discussions.


\bibliographystyle{ieeetr}
\bibliography{ref}
\end{document}